\documentclass[amsmath,amssymb,superscriptaddress,nobalancelastpage,prl,twocolumn]{revtex4-1}
\usepackage{graphics}
\usepackage{bm}
\usepackage{color}
\usepackage{epstopdf}
\usepackage{epsfig}

\begin{document}

\title{Bulk charge stripe order competing with superconductivity in La$_{2-x}$Sr$_x$CuO$_4$ ($x=0.12$)}

\author{N. B. Christensen}
\affiliation{Department of Physics, Technical University of Denmark, DK-2800 Kongens Lyngby, Denmark}
\author{J. Chang} 
\affiliation{Institute for Condensed Matter Physics, \'{E}cole Polytechnique Fed\'{e}rale de Lausanne (EPFL), CH-1015 Lausanne, Switzerland} 
\author{J. Larsen} 
\affiliation{Department of Physics, Technical University of Denmark, DK-2800 Kongens Lyngby, Denmark}
\author{M. Fujita} 
\affiliation{Institute for Materials Research, Tohoku University, Sendai 980-8577, Japan} 
\author{M. Oda}
\affiliation{Department of Physics, Hokkaido University, Sapporo 060-0810, Japan} 
\author{M. Ido}
\affiliation{Department of Physics, Hokkaido University, Sapporo 060-0810, Japan} 
\author{N. Momono} 
\affiliation{Department of Materials Science and Engineering, Muroran Institute of Technology, Muroran, Japan} 
\author{E. M. Forgan} 
\affiliation{School of Physics and Astronomy, University of Birmingham, Birgmingham B15 2TT, United Kingdom}
\author{A. T. Holmes} 
\affiliation{School of Physics and Astronomy, University of Birmingham, Birgmingham B15 2TT, United Kingdom}
\author{J. Mesot} 
\affiliation{Institute for Condensed Matter Physics, \'{E}cole Polytechnique Fed\'{e}rale de Lausanne (EPFL), CH-1015 Lausanne, Switzerland} 
\affiliation{Paul Scherrer Institut, CH-5232 Villigen, Switzerland}
\affiliation{Laboratory for Solid-State Physics, ETH H{\"o}nggerberg, CH-8093 Zurich, Switzerland}
\author{M. Huecker} 
\affiliation{Condensed Matter Physics and Materials Science Department, Brookhaven National Laboratory, Upton, New York 11973, USA} 
\author{M. v. Zimmermann}
\affiliation{Deutsches Elektronen Synchrotron DESY, 22603 Hamburg, Germany}

\begin{abstract}
We present a volume-sensitive high-energy x-ray diffraction study of the underdoped cuprate high temperature superconductor La$_{2-x}$Sr$_x$CuO$_4$ ($x=0.12$, $T_c$=27~K) in applied magnetic field. Bulk short-range charge stripe order with propagation vector ${\bf q}_{\textrm{ch}}=(0.231,~0,~0.5)$ is demonstrated to exist below $T_{\textrm{ch}}=85(10)$~K and shown to compete with superconductivity. We argue that bulk charge ordering arises from fluctuating stripes that become pinned near boundaries between orthorhombic twin domains.
\end{abstract}


\maketitle

A major question in the field of cuprate high-temperature superconductivity concerns the nature of the normal state from which superconductivity emerges in the underdoped region of the phase diagram. Here the opening of a pseudogap manifests itself in the electronic spectrum at temperatures significantly higher than the superconducting transition temperature $T_c$ \cite{Timusk1999}. One school of thought ascribes the pseudogap phenomenon to the emergence of an electronic broken-symmetry phase
competing with superconductivity for the low-energy electronic states.
While ${\bf q}=0$ magnetic order \cite{Fauque_PRL2006, Li_Nature2008, Baledent_PRL2010} remains a candidate pseudogap order parameter, recent attention has focussed on short-range charge density wave (CDW) order. Initially discovered in YBa$_2$CuO$_3$O$_{6+y}$ (YBCO) using x-ray diffraction, and associated with suppressed superconductivity near the hole-doping level $p=0.12$ and evidence for Fermi-surface reconstruction in high magnetic fields \cite{Ghiringhelli2012, Chang2012}, CDW order has subsequently been identified in Bi \cite{Comin2014,SilvaNeto2014,Hashimoto} and Hg-based cuprates \cite{Doiron_PRX2013,Chan_arxiv}. 

In the quest to distinguish universal properties from features specific to particular cuprate families, a crucial outstanding question is the relation between CDW order
\cite{Ghiringhelli2012,Chang2012,Comin2014,SilvaNeto2014,Hashimoto,Doiron_PRX2013,Chan_arxiv,Achkar_PRL2013} and the charge stripe order observed in the La-based cuprates La$_{2-x-y}$Nd$_y$Sr$_x$CuO$_4$ (LNSCO) \cite{Tranquada1995,Zimmermann_1998}, La$_{2-x}$Ba$_x$CuO$_4$ (LBCO) \cite{Fujita2004, Hucker_PRB2011} and La$_{2-x-y}$Eu$_y$Sr$_x$CuO$_4$ (LESCO) \cite{Fink_PRB2009,Hucker_2007}. In all of these materials, where $p=x$, charge order characterized by propagation vectors ${\bf q}_{\textrm{ch}}=(2\delta,~ 0,~ 1/2)$ and $(0,~ 2\delta,~ 1/2)$ is preceded by or coincident with a transition from a low-temperature orthorhombic (LTO, space group Bmab, No. 64) to low-temperature tetragonal or low-temperature-less-orthorhombic (LTT or LTLO; space groups P4$_2/$ncm (138) and Pccn (56), respectively) structures. Quasi two-dimensional incommensurate magnetic order, with propagation vectors ${\bf q}_{\textrm m}=(1/2-\delta,~1/2,~0)$ and $(1/2,~1/2-\delta,~0)$, follows below the charge order transition temperature, $T_{\textrm{ch}}$. Importantly, the in-plane components of ${\bf q}_{{\textrm{m}}}$ and ${\bf q}_{{\textrm{ch}}}$ are intimately related: ${\bf q}_{\textrm{m}}^{\textrm{2D}}=(1/2,~ 1/2) - 1/2~{\bf q}_{{\textrm{ch}}}^{\textrm{2D}}$ \cite{Tranquada1995}. By contrast, in none of the Y, Bi or Hg-based compounds displaying CDW order has an associated incommensurate magnetic order been identified. In fact, the doping-dependence of ${\bf q}_{\textrm{ch}}$ in La-based cuprates is opposite to that of the CDW propagation vector in YBCO, see e.g. \cite{Blackburn2013}. While these observations suggest that CDW and stripe order are distinct electronic instabilities of underdoped cuprates, it has been reported \cite{Hucker_PRB2013} that the charge stripe order in LBCO ($x=0.095$ and $0.155$) responds to temperature and magnetic field in a manner very similar to YBCO \cite{Chang2012,Blackburn2013}. Moreover, albeit x-ray diffraction on orthorhombic La$_{2-x}$Sr$_x$CuO$_4$ (LSCO) has to date only evidenced surface stripe order \cite{Wu_CO}, a recent $^{129}$La NMR study provided indirect evidence for bulk charge stripe order \cite{NMR}. Should charge order be confirmed in the bulk of LSCO, what are its characteristics compared to CDW and stripe ordered cuprates, and what is the role played by structure?

In this Letter, we present direct hard x-ray diffraction evidence for short-range charge stripe order and its competition with superconductivity in  La$_{2-x}$Sr$_x$CuO$_4$ ($x=0.12$). The stripe order is characterized by propagation vector ${\bf q}_{\textrm{ch}}=(0.231,~ 0,~ 1/2)$ and sets in below $T_{\textrm{ch}}=85(10)$~K. We argue that stripe order exists essentially in the bulk of the sample volume but that it is pinned by orthorhombic twin domain boundaries. Our results demonstrate that charge stripe order does not require an average LTT structure and that, once established, it displays exactly the same magnetic field- and temperature dependence as the CDW order in YBCO at a comparable doping level and LBCO at concentrations away from $x=1/8$ \cite{Chang2012,Hucker_PRB2013}. 

\begin{figure}
\begin{center}
\includegraphics[width=0.38\textwidth]{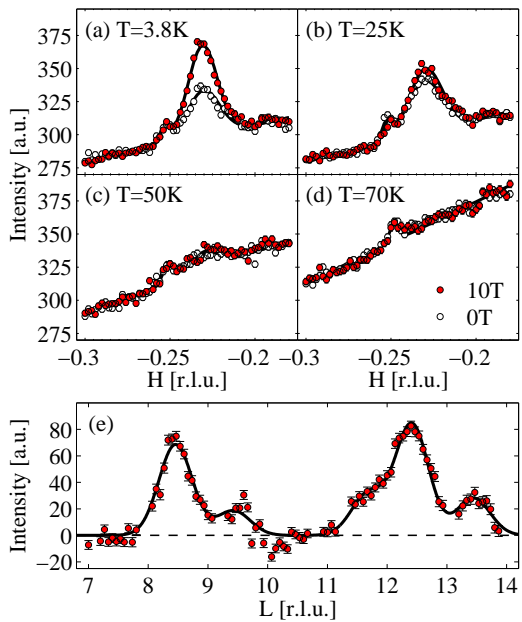}
\end{center}
\caption{Momentum scans through the charge stripe order satellite peak ${\bf Q}_{\textrm {ch}}=(-2\delta,~ 0,~ 8.5)$ in LSCO ($x=0.12$). (a-d) H-scans for temperatures and magnetic fields as indicated. (e) Out-of-plane momentum dependence at 3.8~K and 10~T applied along the $\bf c$-axis. The data in (e) were obtained by scanning along $(0.231,~ 0,~ L)$ and subtracting a background estimate evaluated from the average of scans along $(0.271,~ 0,~ L)$ and $(0.191,~ 0,~ L)$. Regions dominated by powderlines have been removed. All solid lines are fits to Gaussian lineshapes.}
\label{Fig1}
\end{figure}

Our bulk-sensitive hard x-ray (100 keV) diffraction experiments were carried out at the BW5 beamline at DESY, Hamburg, operated in triple-axis mode with Ge-gradient Si(111) monochromator and analyzer crystals and a 2$\times$2 mm aperture to collimate the incident beam. The sample was a 0.8 mm thick platelet with surface normal approximately 45 degrees away from the crystallographic ${\bf a}$ and ${\bf c}$ axes. It was mounted inside a 10~T horizontal field cryomagnet. In transmission geometry, this setup allows access to momentum transfers $(H0L)$ (We employ tetragonal notation with $a \simeq b=3.781$\AA~ and $c=13.169$~\AA), with the field applied either perpendicular or parallel to the CuO$_2$ planes. 

The sample was cut from an image furnace grown single crystal rod. Pieces of the same original rod have been used in previous studies of magnetic and electronic properties of LSCO ($x=0.12$) \cite{Chang2008,Romer_Thesis,Romer_PRB2013,Chang_ARPES,Razzoli_ARPES}. 
Its high-temperature tetragonal (HTT) to LTO transition temperature is $T_{\textrm{LTO}}$=255~K, and superconductivity sets in below $T_{c}$=27~K \cite{Chang2008}. High-resolution elastic neutron scattering identifies incommensurate magnetic Bragg peaks with propagation vector ${\bf q}_{\textrm{m}}=(0.383,~ 1/2,~ 0)$ \cite{Romer_Thesis,Romer_Comment} below $T_\textrm{N} \simeq T_c$ while muon spin rotation, which probes longer time-scale fluctuations, finds evidence for magnetic order only below $T_\mu \simeq$ 11~K \cite{Chang2008,Romer_PRB2013}.

Fig.~\ref{Fig1} illustrates all salient features of our raw data. At the experimental base temperature, 3.8~K (Fig.~\ref{Fig1}(a)), we observe two features on top of a sloping background: A pronounced peak at ${\bf Q}_{\textrm {ch}}=(-0.2306(3),~ 0,~ 8.5)$ corresponding to the propagation vector ${\bf q}_{\textrm {ch}}\simeq (0.231,~ 0,~ 0.5)$ \cite{Propagationvectornote}, and a weak spurious peak near ${\bf Q}_{\textrm {sp}}=(-0.2528(4),~ 0,~ 8.5)$.  The peak at ${\bf Q}_{\textrm {sp}}$ is  not related to charge order since it is not observed at symmetry equivalent wavevectors, and is essentially independent of both temperature and magnetic field. In contrast, 
application of a magnetic field of 10~T at 3.8~K along the crystallographic ${\bf c}$-axis causes a significant enhancement of the ${\bf Q}_{\textrm {ch}}$ peak intensity compared to 
zero field (Fig.~\ref{Fig1}(a)). Just below $T_c$, at 25~K (Fig.~\ref{Fig1}(b)), a peak remains in zero field. This peak is, in fact, more intense than at base temperature, and it is only weakly enhanced by the magnetic field.
At both 50~K and 70~K (Figs.~\ref{Fig1}(c) and (d), respectively) 
the 0~T and 10~T data are identical within errors. In the former case, a broadened profile centered at ${\bf Q}_{\textrm {ch}}$ can be discerned while at 70~K, any remnant signal centered at ${\bf Q}_{\textrm {ch}}$ is very weak. Fig.~\ref{Fig1}(e) illustrates the dependence of the charge order signal on the out-of-plane momentum direction. Prominent broad peaks are observed near half-integer $L$. 

\begin{figure}
\begin{center}
\includegraphics[width=0.49\textwidth]{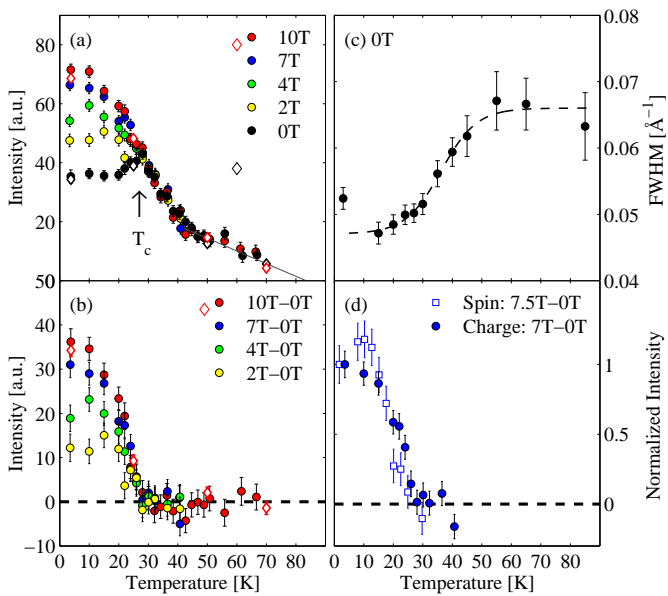}
\end{center}
\caption{(a) Temperature and magnetic field dependence of the raw peak intensity at $(-0.231,~ 0,~ 8.5)$ (see text for details) (b) Field-induced charge order intensity, determined by subtracting the zero-field data from  finite field data. (c) Temperature dependence of the charge order peak full-width-half-maximum for $\mu_0H=0$~T. (d) Comparison of the field-induced charge order intensity at $\mu_0H=7$~T to the field-induced magnetic intensity measured at ${\bf q}_{\textrm {m}}$ and $\mu_0H=7.5$~T (Produced using raw data from Ref. \cite{Chang2008}). Both quantities have been normalized to their value at base temperature.}
\label{Fig2}
\end{figure}

We emphasize that the components of the charge and spin propagation vectors ${\bf q}_{\textrm {ch}}=(-0.231,~ 0,~ 1/2)$ and ${\bf q}_{\textrm {m}}=(0.383,~ 1/2,~ 0)$ \cite{Romer_Thesis,Romer_Comment} parallel to the CuO$_2$ planes obey the relation ${\bf q}_{\textrm {m}}^{2D}=(1/2,~ 1/2) \pm 1/2~{\bf q}_{\textrm {ch}}^{2D}$ common to charge-stripe ordered materials \cite{Tranquada1995,Fujita2004,Fink_PRB2009}.
Thus, we conclude that the observed ${\bf q}_{\textrm {ch}}$ peak in LSCO ($x=$0.12) reflects the lattice response to charge stripe ordering in orthorhombic LSCO (See Ref. \cite{Thampy2014} for further support of stripe order in our crystal). 

Fig.~\ref{Fig2}(a) illustrates the temperature and field-dependence of the charge stripe order signal in more detail. To efficiently probe these dependencies we subtracted a background estimate determined as the average of the intensities at $(-0.191,~ 0,~ 8.5)$ and $(-0.271,~ 0,~ 8.5)$ from the peak intensity collected at $(-0.231,~ 0,~ 8.5)$ (closed circles).  This procedure is justified by the fact that the background in Figs.~\ref{Fig1}(a)-(d) is linear to a good approximation. Analysis of full scans (open diamonds in Fig.~\ref{Fig2}(a)) yield essentially identical results. For $\mu_0H=0$T, the charge stripe peak emerges gradually below an onset temperature which by extrapolation we estimate to be $T_{\textrm {ch}}~85(10)$~K (grey line in Fig.~\ref{Fig2}(a)). Upon cooling the peak becomes more intense until $T \simeq T_c$.  Strikingly, for $T < T_c$, a distinct reduction of the intensity is seen, clearly reflecting the competition between charge order and bulk superconductivity. This intensity suppression strongly resembles 
the CDW response in YBCO \cite{Ghiringhelli2012,Chang2012,Blackburn2013} and the charge stripe response in LBCO away from $p=1/8$ \cite{Hucker_PRB2013}. 

\begin{figure}
\begin{center}
\includegraphics[width=0.45\textwidth]{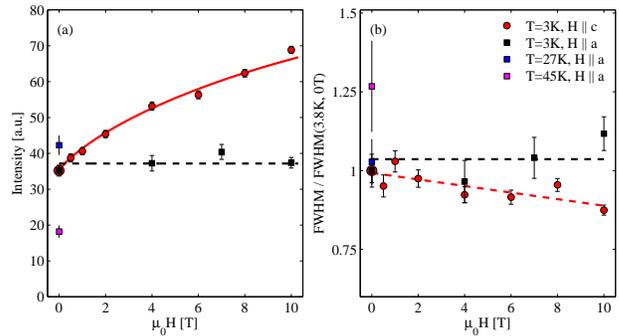}
\end{center}
\caption{(a) Magnetic field dependence of the peak intensity and (b) normalized peak full-width-half-maximum for fields along the $\bf c$-axis (circles; ${\bf Q}_{\textrm{ch}}=(-0.231,~ 0,~ 8.5)$) and $\bf a$-axis (squares; ${\bf Q}_{\textrm{ch}}=(-0.231,~ 0,~ 12.5)$). All lines are guides to the eye. The solid red line in (a) is based on the functional form $I(H)=I_0 + I_1(H/H_{c2})\log({H_{c2}/H})$ \cite{Demler2001}.}
\label{Fig3}
\end{figure}

Application of a magnetic field along the ${\bf c}$-axis suppresses superconductivity and, for temperatures smaller than $T_c$ only, enhances the charge stripe order intensity.  This last point is brought out most clearly in Fig.~\ref{Fig2}(b) where we plot the difference between peak intensities measured in finite field and zero field. The enhancement of the charge stripe peak intensity is clear already at $2$~T and continues to the highest field probed.  The field-dependence at base temperature, extracted from fits to full momentum scans, similar to those shown in Fig.~\ref{Fig1}(a), is plotted in Fig.~\ref{Fig3}(a). In contrast, the charge stripe signal is unaffected by a magnetic field applied along ${\bf a}$. This anisotropy parallels pioneering neutron scattering studies of field-induced magnetic order in underdoped LSCO \cite{Lake2002,Lake2005} and reflects the inefficiency of in-plane fields in suppressing bulk superconductivity ($H_{c2,ab} \gg H_{c2,c}$). Fig.~\ref{Fig2}(d) highlights how the charge and spin components of the stripe order rise in tandem when superconductivity is suppressed, demonstrating that the two are closely linked.

The characteristic temperature and field-dependencies of the satellite peak intensity (Figs.~\ref{Fig2}(a) and \ref{Fig3}(a)) appear to be universal signatures of microscopically coexisting but competing superconducting and charge order parameters
\cite{Ghiringhelli2012,Chang2012,Hucker_PRB2013}, irrespective of the stripe or density wave nature of the latter. A further hallmark of charge ordered cuprates is broader peaks along the $\bf c$-axis than along in-plane directions. We can quantify this anisotropy for LSCO ($x=$0.12) by fitting Gaussian lineshapes to data such as those shown in Fig.~\ref{Fig1} (solid lines) and extracting correlation lengths, $\xi_a$ and $\xi_c$ as the inverse half-width-at-half-maximum \cite{Resolution_Note}. While this always yielded $\xi_c \sim 6.3$ \AA~, implying that the charge stripe order is uncorrelated beyond nearest neighbor CuO$_2$ layers, the in-plane correlation length, $\xi_a$, was found to vary between different experiments. The data in Figs.~\ref{Fig1}(a)-(d) yield a charge stripe correlation length $\xi_a =53$ \AA~ at 3.8~K and zero field, while a subsequent experiment performed under identical conditions gave $\xi_a =38$ \AA. This suggests that the charge stripe order is not completely homogeneous in our sample. As will be discussed below, this seems to have a physical reason, and is not related to the crystal's (high) quality.
The key observation, however, is that the evolution of the in-plane correlation length with temperature and magnetic field did not vary between experiments even if the absolute values did. Figs. \ref{Fig2}(b) and \ref{Fig3}(a) show, respectively, the temperature dependence of the peak width and the field-dependence of the same quantity, normalized to its zero-field, base temperature value. Upon cooling, the peak gradually sharpens up, reaching a mimimum below $T_c$ and displaying a small upturn at the lowest temperatures. Similar behaviour has been reported for YBCO \cite{Ghiringhelli2012,Chang2012}. As a function of magnetic field along the $\bf c$-axis, a $\sim$ 10\%~ sharpening, reflecting a slight increase of the charge order correlation length, was observed, while an in-plane field leaves the peak width unchanged.

\begin{figure}
\begin{center}
\includegraphics[width=0.48\textwidth]{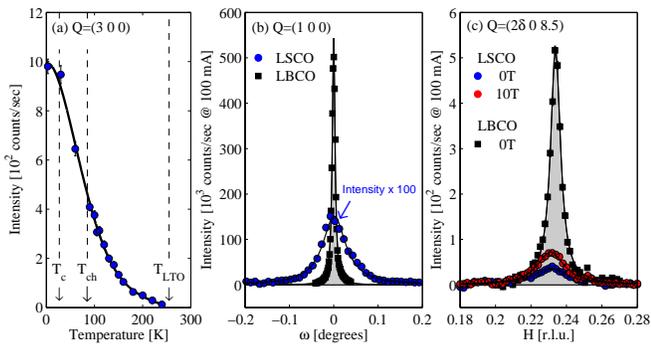}
\end{center}
\caption{(a) Temperature dependence of the $(300)$ Bragg reflection. Dashed lines indicate $T_{c}$, $T_{\textrm{ch}}$ and $T_{\textrm {LTO}}$. The full line is a guide to the eye. (b) and (c):  Comparisons of $(100)$ and $(2\delta,~0,~8.5)$ peaks in LSCO $p=0.12$ (red and blue circles) and LBCO $p=1/8$ (squares) \cite{Hucker_PRB2011}, respectively. Lines in (b) and (c) are fits performed to evalutate the integrated intensities. Note the different intensity scales in (b).}
\label{Fig4}
\end{figure}

Finally, we turn to the relation between charge order and the crystal structure of LSCO. Fig.~\ref{Fig4}(a) shows the temperature dependence of a weak signal observed at $(300)$. Peaks of the type $(H~0~0)$ with $H={\textrm{odd}}$ are structurally forbidden in the LTO phase but permitted when the symmetry is lowered to LTT or LTLO \cite{Axe,Tranquada1995,Fujita2004,Fink_PRB2009}.  Three experimental facts, however, allow us to argue that the peak in Fig. \ref{Fig4}(a) is not the signature of an average LTT or LTLO structure in our sample: (i) The LTO $\rightarrow$ LTT/LTLO transition is first order (see e.g  Ref. \cite{Hucker_PRB2011}), whereas the $(300)$ peak emerges gradually with cooling. (ii) The peak appears just below the known HTT $\rightarrow$ LTO transition temperature $T_{\textrm{LTO}}=255$~K \cite{Chang2008} of our sample. (iii) The integrated intensity of the $(100)$ peak is weaker by a factor of 42 compared to the intrinsic LTT phase of LBCO at $p=1/8$ (Fig. \ref{Fig4}(b)), measured under identical conditions \cite{Hucker_PRB2011} and normalized by flux and sample volume, while the $0$~T and $10$~T charge order peaks LSCO are weaker only by factors of 4.2 and 2.6, respectively, than the fully saturated stripe order in LBCO, p=$1/8$ \cite{Hucker_PRB2013} (Fig. \ref{Fig4}(c)). This shows that the charge stripe peak in LSCO is substantially smaller than the same peak in LBCO, but not as small as expected in the case of a linear scaling with the $(100)$ peak intensity.

These considerations led us to conclude that the $(100)$ and $(300)$ peaks must be related to x-rays diffracted by orthorhombic twin domain boundaries. From transmission electron microscopy studies of samples with compositions similar to ours \cite{Horibe1,Horibe2} it is well known that these boundaries between LTO twin domains exhibit LTT-type reflections (See Ref. \cite{Braden1992} for a atomic level explanation). Consideration of the known stripe-stabilizing potential of LTT or LTLO structures \cite{Tranquada1995,Fujita2004,Fink_PRB2009} then suggests that structural defects in the form of domain boundaries with local LTT or LTLO structures cause the emergence of charge order by pinning low-energy fluctuating stripes \cite{Kivelson2003}. On the other hand, to explain the relatively large intensity ratio of ${\bf q}_{\textrm{ch}}$ to $(300)$ peaks, charge stripe order must extend significantly beyond the pinning centers and into parts of the sample displaying the average LTO structure. 

This scenario implies that stripe order in orthorhombic LSCO most likely depends not only on doping, reflecting a bare charge order susceptibility, and potential commensurability effects, but also on sample-specific details such as the orthorhombic twin domain structure. Charge stripe inhomogeneity and glassiness \cite{NMR,Mitrovic_PRB2008} are obvious possible consequences. The above considerations may explain why, in a crystal with $x=0.10$ that shows no evidence for LTT-type reflections, we also found no charge order. Furthermore, potential differences of the domain structure of the surface and bulk of a single crystal may explain why the x-ray study in Ref. \cite{Wu_CO} found charge stripe ordering near the surface, but not in the bulk, of a sample of nominally the same composition as ours. Indeed surface regions with local LTT/LTLO structure may be implicated in observations of shadow-bands in angle resolved photoemission studies of LSCO \cite{Zhou_ARPES,Chang_ARPES,Razzoli_ARPES,He_ARPES}. We also note that in samples with average LTT structure, such as LBCO, the  direction of stripes alternate between the $\bf a$ and $\bf b$ axes, causing charge peaks to center at half-integer $L$ \cite{Tranquada1995,Zimmermann_1998}. There is no {\it a priori} reason for this to occur in the LTO structure. The fact that the peaks in Fig.~\ref{Fig1}(e) do occur at approximately half-integer $L$ therefore strengthens the case for stripe pinning near twin domain boundaries with local LTT/LTLO structure. 

Our direct detection of charge order in the bulk of LSCO brings understanding to a host of experiments of the past two decades (See e.g. Refs. \cite{Kumagai_PhysicaB,Radaelli_PRB1994,Goto_JPSJ,Julien_PRB2001,Suzuki_PRB1998,Watanabe_Hyperfine}), which suggested that the weak dip in the superconducting $T_c$-dome of LSCO \cite{Takagi_1989} may be related to the much more pronounced $1/8$-anomaly in LBCO \cite{Moodenbaugh} and to charge stripe order, but fell short of providing direct evidence. Similarly, in the light of our results, giant phonon anomalies observed in superconducting LSCO near ${\bf q}_{\textrm{ch}}$ and indirectly associated with collective charge excitations (See e.g. Ref. \cite{Park_PRB2014}), can now be directly related to incipient charge stripe order.

In conclusion, this Letter has demonstrated the existence of charge stripe order competing with superconductivity in the bulk of La$_{2-x}$Sr$_x$CuO$_4$ ($x=0.12$). 
Our crystal structure analysis and intensity comparison implies that the charge stripe order is pinned by tetragonal twin domain boundaries of the orthorhombic parent phase but extends far beyond these and into the orthorhombic bulk. 

This work was supported the Danish Agency for Science, Technology, and Innovation under
DANSCATT and the Swiss National Science Foundation through NCCR-MaNEP and Grant No. PZ00P2$_{}$142434. Work at Brookhaven was supported by the U.S. Department of Energy (DOE), under Contract No. DE-AC02-98CH10886.


\begin{thebibliography}{99}

\bibitem{Timusk1999} T. Timusk, and B. Statt, Rep. Prog. Phys. {\bf 62}, 61 (1999).

\bibitem{Fauque_PRL2006} B. Fauqu\'e, Y. Sidis, V. Hinkov, S. Pailh`es, C. T. Lin, X. Chaud, and P. Bourges, Phys. Rev. Lett. {\bf 96}, 197001 (2006).

\bibitem{Li_Nature2008} Y. Li, V. Bal{\'e}dent, N. Bari{\v{s}}i{\'c}, Y. Cho, B. Fauqu{\'e}, Y. Sidis, G. Yu, X. Zhao, P. Bourges, and M. Greven, Nature {\bf 455}, 372 (2008).

\bibitem{Baledent_PRL2010} V. Bal\'edent, B. Fauqu\'e, Y. Sidis, N. B. Christensen, S. Pailh{\`e}s, K. Conder, E. Pomjakushina, J. Mesot, and P. Bourges, Phys. Rev. Lett. {\bf 105}, 027004 (2010).

\bibitem{Ghiringhelli2012} G. Ghiringhelli {\it{et al.}}, Science, {\bf 337}, 821 (2012).

\bibitem{Chang2012} J. Chang {\it{et al.}}, Nature Physics, {\bf 8}, 871 (2012).

\bibitem{Comin2014} R. Comin {\it{et al.}}, Science {\bf 343}, 390 (2014).

\bibitem{SilvaNeto2014} E. H. d. Silva Neto {\it{et al.}}, Science {\bf 343}, 393 (2014).

\bibitem{Hashimoto} M. Hashimoto {\it{et al.}}, arXiv:1403.0061

\bibitem{Doiron_PRX2013} N. Doiron-Leyraud {\it{et al.}}, Phys. Rev. X {\bf 3}, 021019 (2013).

\bibitem{Chan_arxiv} M. K. Chan {\it{et al.}}, arXiv:1402.4517

\bibitem{Achkar_PRL2013} A. J. Achkar {\it{et al.}}, Phys. Rev. Lett. {\bf 109}, 167001 (2013).

\bibitem{Tranquada1995} J. M. Tranquada, B. J. Sternlieb, J. D. Axe, Y. Nakamura, and S. Uchida, Nature (London) {\bf 375}, 561 (1995).

\bibitem{Zimmermann_1998} M. v. Zimmermann {\it{et al.}}, Europhys. Lett. {\bf 41}, 629 (1998). 

\bibitem{Fujita2004} M. Fujita, H.Goka, K.Yamada, J. M. Tranquada, and L. P. Regnault, Phys. Rev. B {\bf 70}, 104517 (2004).

\bibitem{Hucker_PRB2011} M. H{\"u}cker, M. v. Zimmermann, G. D. Gu, Z. J. Xu, J. S. Wen, G. Xu, H. J. Kang, A. Zheludev and J. M. Tranquada, Phys. Rev. B {\bf 83}, 104506 (2011). 

\bibitem{Fink_PRB2009} J. Fink {\it{et al.}}, Phys. Rev. B {\bf 79}, 100502(R) (2009).

\bibitem{Hucker_2007} M. H{\"u}cker, G.D. Gu, J.M. Tranquada, M.v. Zimmermann, H.-H. Klauss, N.J. Curro, M. Braden, and B. B{\"u}chner, Physica C ${\textbf {460}}-{\textbf {462}}$, 170–173 (2007). 

\bibitem{Blackburn2013} E. Blackburn {\it{et al.}}, Phys. Rev. Lett. {\bf 110}, 137004 (2013).

\bibitem{Hucker_PRB2013} M. H{\"u}cker, M. v. Zimmermann, Z. J. Xu, J. S. Wen, G. D. Gu, and J. M. Tranquada, Phys. Rev. B {\bf 87}, 014501 (2013).

\bibitem{Wu_CO} H.-H. Wu {\it{et al.}}, Nat. Commun. 3:1023 doi: 10.1038/ncomms2019 (2012).

\bibitem{NMR} S.-H. Baek, M. H{\"u}cker, A. Erb, G. D. Gu, B. B{\"u}chner and H.-J. Grafe, arXiv:1402.3077

\bibitem{Chang2008} J. Chang {\it{et al.}}, Phys. Rev. B {\bf 78}, 104525 (2008).

\bibitem{Romer_Thesis} A. T. R{\o}mer, {\it Magnetic correlations in the high-temperature
superconductor La$_{1.88}$Sr$_{0.12}$CuO$_4$}, Master Thesis, University of Copenhagen (2009). 

\bibitem{Romer_PRB2013} A. T. R{\o}mer {\it{et al.}}, Phys. Rev. B {\bf 87}, 144513 (2013).

\bibitem{Chang_ARPES} J. Chang {\it{et al.}}, New J. Phys. {\bf 10}, 103016 (2008).

\bibitem{Razzoli_ARPES} E. Razzoli {\it{et al.}}, New J. Phys. {\bf 12}, 125003 (2010).

\bibitem{Romer_Comment} With respect to the main structural twin domain, incommensurate magnetic reflections were identified at $(−0.1240,~ 1.1022,~ 0)$ and $(−1.1244,~ 0.1149,~ 0)$ in orthorhombic notation \cite{Romer_Thesis}. Converting to the pseudotetragonal setting used in this Letter, the corresponding average magnetic propagation vector is ${\bf q}_m=(0.5-\delta,~ 0.5,~ 0)$ with $\delta \simeq 0.117$. The deviation of the peak positions from symmetry directions is similar to the observations reported in Ref. \cite{Suzuki}.

\bibitem{Suzuki} H. Kimura, H. Matsushita, K. Hirota, Y. Endoh, K. Yamada, G. Shirane, Y. S. Lee, M. A. Kastner, and R. J. Birgeneau, Phys. Rev. B {\bf 61}, 14366 (2000).

\bibitem{Propagationvectornote} The propagation vector ${\bf q}_{\textrm{ch}}$ is defined in the first Brillouin zone, while ${\bf Q}_{\textrm{ch}}$ denotes a general satellite peak position. The two are related by ${\bf Q}_{\textrm{ch}}={\bf G} \pm {\bf q}_{\textrm{ch}}$, where $\bf G$ is a reciprocal lattice vector.

\bibitem{Thampy2014} V. Thampy {\it{et al.}}, Manuscript under preparation.

\bibitem{Demler2001} E. Demler, S. Sachdev, and Y. Zhang, Phys. Rev. Lett. {\bf 87}, 067202 (2001).

\bibitem{Lake2002} B. Lake {\it{et al.}}, Nature {\bf 415}, 299 (2002).

\bibitem{Lake2005} B. Lake {\it{et al.}}, Nature Materials {\bf 4}, 658 (2005). 

\bibitem{Resolution_Note} Because our experimental resolution is significantly worse along the $\bf b$-axis, we limit our discussion to the correlation lengths along $\bf a$ and $\bf c$ and refer the reader to Ref. \cite{Thampy2014} for important details regarding the in-plane behaviour of the charge order in a piece of the same original sample.

\bibitem{Axe} J. D. Axe, A. H. Moudden, D. Hohlwein, D. E. Cox, M. Mohanty, A. R. Moodenbaugh, and Y. Xu, Phys. Rev. Lett. {\bf 62}, 2751 (1989).

\bibitem{Horibe1} Y. Horibe, Y. Inoue, and Y. Koyama, Physica C ${\textbf {282}}-{\textbf {287}}$, 1071 (1997).

\bibitem{Horibe2} Y. Horibe, Y. Inoue, and Y. Koyama, Phys. Rev. B {\bf 61}, 11 922 (2000).

\bibitem{Braden1992} M. Braden, G. Heger, P. Schweiss, Z. Fisk, K. Gamayunov, I. Tanaka, and H. Kojima, Physica C {\bf 191}, 455 (1992).

\bibitem{Kivelson2003} S. A. Kivelson, I. P. Bindloss, E. Fradkin, V. Oganesyan, J. M. Tranquada, A. Kapitulnik, and C. Howald, Rev. Mod. Phys. {\bf 75}, 1201 (2003).

\bibitem{Mitrovic_PRB2008} V. F. Mitrovi{\'c}, M.-H. Julien, C. de Vaulx, M. Horvati{\'c}, C. Berthier, T. Suzuki, and K. Yamada, Phys. Rev. B {\bf 78}, 014504 (2008).

\bibitem{Zhou_ARPES} X. J. Zhou {\it{et al.}}, Phys. Rev. Lett. {\bf 92}, 187001 (2004).

\bibitem{He_ARPES} R.-H. He {\it{et al.}}, New J. Phys. {\bf 13}, 013031 (2011).

\bibitem{Kumagai_PhysicaB} K. Kumagai K. Kawano, H. Kagami, G. Suzuki, Y. Matsuda, I. Watanabe, K. Nishiyama, and K. Nagamine, Physica C ${\textbf {235}}-{\textbf {240}}$, 1715 (1994).

\bibitem{Radaelli_PRB1994} P. G. Radaelli, D. G. Hinks, A. W. Mitchell, B. A. Hunter, J. L. Wagner, B. Dabrowski, K. G. Vandervoort, H. K. Viswanathan, and J. D. Jorgensen, Phys. Rev. B {\bf 49}, 4163 (1994).

\bibitem{Goto_JPSJ} T. Goto, K. Chiba, M. Mori, T. Suzuki, K. Seki, and T. Fukase, J. Phys. Soc. Japan {\bf 66}, 2870 (1997).

\bibitem{Julien_PRB2001} M.-H. Julien {\it{et al.}}, Phys. Rev. B {\bf 63}, 144508 (2001).

\bibitem{Suzuki_PRB1998} T. Suzuki, T. Goto, K. Chiba, T. Shinoda, T. Fukase, H. Kimura, K. Yamada, M. Ohashi and Y. Yamaguchi, Phys. Rev. B {\bf 57}, R3229(R) (1998).

\bibitem{Watanabe_Hyperfine} I. Watanabe, K. Nishiyama, K. Nagamine, K. Kawano, and K. Kumagai, Hyperfine Interactions {\bf 86}, 603 (1995).

\bibitem{Takagi_1989} H. Takagi, T. Ido, S. Ishibashi, M. Uota, S. Uchida, Y. Tokura, Phys. Rev. B {\bf 40}, 2254 (1989).

\bibitem{Moodenbaugh} A. R. Moodenbaugh, Y. Xu, M. Suenaga, T. J. Folkerts, and R. N. Shelton, Phys. Rev. B {\bf 38}, 4596 (1988).

\bibitem{Park_PRB2014} S. R. Park, T. Fukuda, A. Hamann, D. Lamago, L. Pintschovius, M. Fujita, K. Yamada, and D. Reznik, Phys. Rev. B {\bf 89}, 020506(R) (2014). 

\end{thebibliography}
\end{document}